%% file: ice_paper.tex
\documentclass[12pt]{article}

\include{preamble}

\newtoggle{usejpgs}
\toggletrue{usejpgs}

\title{\bf Peeking Inside the Black Box: Visualizing Statistical Learning with Plots of Individual Conditional Expectation}
\author{Alex Goldstein\thanks{Electronic address: \texttt{alexg@wharton.upenn.edu}; Principal Corresponding author}~}
\author{Adam Kapelner\thanks{Electronic address: \texttt{kapelner@wharton.upenn.edu}; Corresponding author}}
\author{Justin Bleich\thanks{Electronic address: \texttt{jbleich@wharton.upenn.edu}; Corresponding author}}
\author{Emil Pitkin\thanks{Electronic address: \texttt{pitkin@wharton.upenn.edu}; Corresponding author}}
\affil{The Wharton School of the University of Pennsylvania}

\begin{document}
\maketitle

\begin{abstract}
This article presents Individual Conditional Expectation (ICE) plots, a tool for visualizing the model estimated by any supervised learning algorithm.  Classical partial dependence plots (PDPs) help visualize the average partial relationship between the predicted response and one or more features. In the presence of substantial interaction effects, the partial response relationship can be heterogeneous. Thus, an average curve, such as the PDP, can obfuscate the complexity of the modeled relationship. Accordingly, ICE plots refine the partial dependence plot by graphing the functional relationship between the predicted response and the feature for \textit{individual} observations. Specifically, ICE plots highlight the variation in the fitted values across the range of a covariate, suggesting where and to what extent heterogeneities might exist. In addition to providing a plotting suite for exploratory analysis, we include a visual test for additive structure in the data generating model. Through simulated examples and real data sets, we demonstrate how ICE plots can shed light on estimated models in ways PDPs cannot. Procedures outlined are available in the \texttt{R} package \texttt{ICEbox}.
\end{abstract}

\section{Introduction}\label{sec:introduction}

The goal of this article is to present Individual Conditional Expectation (ICE) plots, a toolbox for visualizing models produced by \qu{black box} algorithms.  These algorithms use training data  $\{\x_i,\,y_i\}_{i=1}^{N}$ (where $\x_i = (x_{i,1}, \ldots, x_{i,p})$ is a vector of predictors and $y_i$ is the response) to construct a model $\hat{f}$ that maps the features $\x$ to fitted values $\hat{f}(\x)$. Though these algorithms can produce fitted values that enjoy low generalization error, it is often difficult to understand how the resultant $\hat{f}$ uses $\x$ to generate predictions.  The ICE toolbox helps visualize this mapping.

ICE plots extend \citet{Friedmana}'s Partial Dependence Plot (PDP), which highlights the average partial relationship between a set of predictors and the predicted response. ICE plots disaggregate this average by displaying the estimated functional relationship for each observation.  Plotting a curve for each observation helps identify interactions in $\hat{f}$ as well as extrapolations in predictor space.

The paper proceeds as follows. Section~\ref{sec:background} gives background on visualization in machine learning and introduces PDPs more formally.  Section~\ref{sec:algorithm} describes the procedure for generating ICE plots and its associated plots. In Section~\ref{sec:simulations} simulated data examples illustrate that ICE plots can be used to identify features of $\hat{f}$ that are not visible in PDPs, or where the PDPs may even be misleading.  Each example is chosen to illustrate a particular principle.  Section~\ref{sec:real_data} provides examples of ICE plots on real data. In Section \ref{sec:testing} we shift the focus from the fitted $\hat{f}$ to a data generating process $f$ and use ICE plots as part of a visual test for additivity in $f$. Section~\ref{sec:discussion} concludes.

\section{Background}\label{sec:background}

\subsection{Survey of Black Box Visualization}

There is an extensive literature that attests to the superiority of black box machine learning algorithms in minimizing predictive error, both from a theoretical and an applied perspective. \citet{Breiman2001}, summarizing, states \qu{accuracy generally requires more complex prediction methods ...[and] simple and interpretable functions do not make the most accurate predictors.} Problematically, black box models offer little in the way of interpretability, unless the data is of very low dimension. When we are willing to compromise interpretability for improved predictive accuracy, any window into black box's internals can be beneficial. 


Authors have devised a variety of algorithm-specific techniques targeted at improving the interpretability of a particular statistical learning procedure's output.  \citet{Rao1997} offers a technique for visualizing the decision boundary produced by bagging decision trees. Although applicable to high dimensional settings, their work primarily focuses on the low dimensional case of two covariates. \citet{Tzeng2005} develops visualization of the layers of neural networks to understand dependencies between the inputs and model outputs and yields insight into classification uncertainty. \citet{Jakulin2005} improves the interpretability of support vector machines by using a device called \qu{nomograms} which provide graphical representation of the contribution of variables to the model fit.  Pre-specified interaction effects of interest can be displayed in the nomograms as well.  \cite{Breiman2001a} uses randomization of out-of-bag observations to compute a variable importance metric for Random Forests (\texttt{RF}). Those variables for which predictive performance degrades the most vis-a-vis the original model are considered the strongest contributors to forecasting accuracy. This method is also applicable to stochastic gradient boosting \citep{Friedman2002}.  \citet{Plate2000} plots neural network predictions in a scatterplot for each variable by sampling points from covariate space. Amongst the existing literature, this work is the most similar to ICE, but was only applied to neural networks and does not have a readily available implementation. 

Other visualization proposals are model agnostic and can be applied to a host of supervised learning procedures.  For instance, \citet{Strumbelj2011} consider a game-theoretic approach to assess the contributions of different features to predictions that relies on an efficient approximation of the Shapley value. \cite{Jiang2002} use quasi-regression estimation of black box functions. Here, the function is expanded into an orthonormal basis of coefficients which are approximated via Monte Carlo simulation. These estimated coefficients can then be used to determine which covariates influence the function and whether any interactions exist. 

\subsection{Friedman's PDP}

Another particularly useful model agnostic tool is \citet{Friedmana}'s PDP, which this paper extends.  The PDP plots the change in the average predicted value as specified feature(s) vary over their marginal distribution.  Many supervised learning models applied across a number of disciplines have been better understood thanks to PDPs. \citet{Green2010a} use PDPs to understand the relationship between predictors and the conditional average treatment effect for a voter mobilization experiment, with the predictions being made by Bayesian Additive Regression Trees \citep[\texttt{BART},][]{Chipman2010a}. \citet{Berk2013} demonstrate the advantage of using \texttt{RF} and the associated PDPs to accurately model predictor-response relationships under asymmetric classification costs that often arise in criminal justice settings. In the ecological literature, \citet{Elith2008a}, who rely on stochastic gradient boosting, use PDPs to understand how different environmental factors influence the distribution of a particular freshwater eel. 

To formally define the PDP, let $S\subset \{1,...,p\}$ and let $C$ be the complement set of $S$.  Here $S$ and $C$ index subsets of predictors; for example, if $S=\{1,2,3\}$, then $\x_S$ refers to a $3 \times 1$ vector containing the values of the first three coordinates of $\x$.  Then the partial dependence function of $f$ on $\x_S$ is given by 

\bneqn\label{eq:true_pdp}
f_S = \expesub{\x_C}{f(\x_S,\x_C)} = \int f(\x_S,\x_C) \mathrm{dP}\parens{\x_C}.
\eneqn

Each subset of predictors $S$ has its own partial dependence function $f_S$, which gives the average value of $f$ when $\x_S$ is fixed and $\x_C$ varies over its marginal distribution $\mathrm{dP}\parens{\x_C}$. As neither the true $f$ nor $\mathrm{dP}\parens{\x_C}$ are known, we estimate Equation \ref{eq:true_pdp} by computing

\bneqn\label{est_pdp}
\hat{f}_S = \frac{1}{N}\sum\limits_{i=1}^N \hat{f}(\x_S,\x_{Ci})
\eneqn

\noindent where $\{\x_{C1},...,\x_{CN}\}$ represent the different values of $\x_C$ that are observed in the training data.  Note that the approximation here is twofold: we estimate the true model with $\hat{f}$, the output of a statistical learning algorithm, and we estimate the integral over $\x_C$ by averaging over the $N$ $\x_C$ values observed in the training set.

This is a visualization tool in the following sense: if $\hat{f}_S$ is evaluated at the $\x_S$ observed in the data, a set of $N$ ordered pairs will result: $\{(\x_{S\ell}, \hat{f}_{S\ell})\}_{\ell=1}^{N}$, where $\hat{f}_{S\ell}$ refers to the estimated partial dependence function evaluated at the $\ell$th coordinate of $\x_S$, denoted $\x_{S\ell}$.  Then for one or two dimensional $\x_S$, \citet{Friedmana} proposes plotting the $N$ $\x_{S\ell}$'s versus their associated $\hat{f}_{S\ell}$'s, conventionally joined by lines.  The resulting graphic, which is called a partial dependence plot, displays the average value of $\hat{f}$ as a function of $\x_S$.  For the remainder of the paper we consider a single predictor of interest at a time ($|S|=1$) and write $x_S$ without boldface accordingly.

As an extended example, consider the following data generating process with a simple interaction:

\bneqn\label{eq:criss_cross_model0}
&& Y = 0.2 X_{1} - 5X_{2}  + 10X_{2} \indic{X_{3} \geq 0} + \errorrv, \\ \nonumber
&& \errorrv \iid \normnot{0}{1}, \quad X_{1}, X_{2}, X_{3} \iid \uniform{-1}{1}.
\eneqn

\begin{figure}[htp]
\centering
\begin{subfigure}[b]{0.48\textwidth}
                \centering
                \includegraphics[width=2.35in]{criss_cross_x2_vy.\iftoggle{usejpgs}{jpg}{\iftoggle{usejpgs}{jpg}{pdf}}}
                \caption{Scatterplot of $Y$ versus $X_2$}
                \label{fig:criss_cross_x2vy}
        \end{subfigure}~~
\begin{subfigure}[b]{0.48\textwidth}
                \centering
                \includegraphics[width=2.35in]{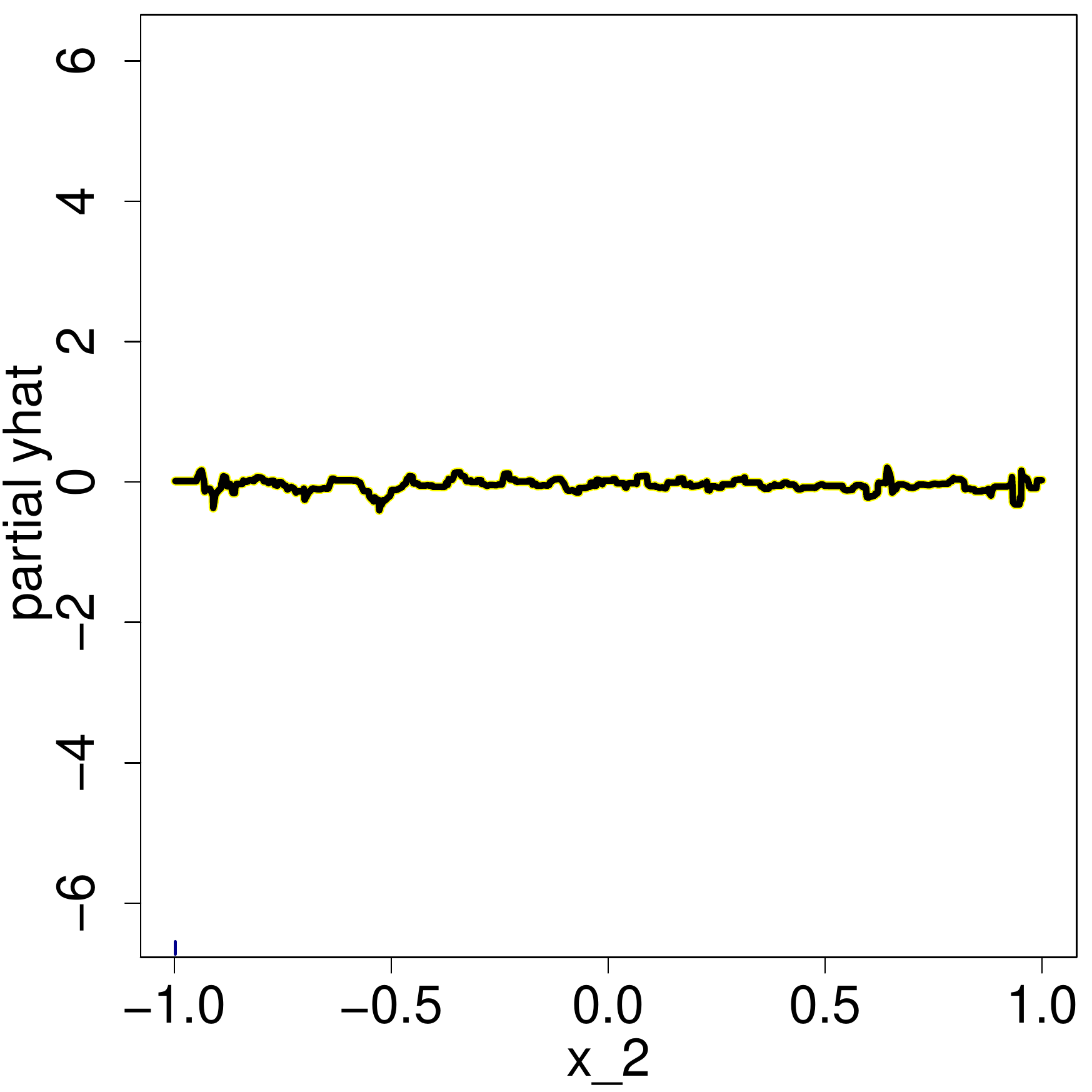}
                \caption{PDP}
                \label{fig:criss_cross_x2_pdp}
        \end{subfigure}
\caption{Scatterplot and PDP of $X_2$ versus $Y$ for a sample of size 1000 from the process described in Equation \ref{eq:criss_cross_model0}. In this example $\hat{f}$ is fit using \texttt{SGB}. The PDP incorrectly suggests that there is no meaningful relationship between $X_2$ and the predicted $Y$.}
\label{fig:no_inter_example}
\end{figure}

We generate 1,000 observations from this model and fit a stochastic gradient boosting model (\texttt{SGB}) via the \texttt{R} package \texttt{gbm} \citep{Ridgeway2013} where the number of trees is chosen via cross-validation and the interaction depth is set to 3.  We now consider the association between predicted $Y$ values and $X_2$ ($S=X_2)$. In Figure \ref{fig:criss_cross_x2vy} we plot $X_2$ versus $Y$ in our sample.  Figure \ref{fig:criss_cross_x2_pdp} displays the fitted model's partial dependence plot for predictor $X_2$. The PDP suggests that on average, $X_2$ is not meaningfully associated with the predicted $Y$.  In light of Figure \ref{fig:criss_cross_x2vy}, this conclusion is plainly wrong. Clearly $X_2$ is associated with $Y$; it is simply that the averaging inherent in the PDP shields this discovery from view.

In fact, the original work introducing PDPs argues that the PDP can be a useful summary for the chosen subset of variables if their dependence on the remaining features is not too strong. When the dependence is strong, however -- that is, when interactions are present -- the PDP can be misleading. Nor is the PDP particularly effective at revealing extrapolations in $\mathcal{X}$-space. ICE plots are intended to address these issues.

\section{The ICE Toolbox}\label{sec:algorithm}

\subsection{The ICE Procedure}\label{subsec:vanilla_plot}

Visually, ICE plots disaggregate the output of classical PDPs.  Rather than plot the target covariates' \textit{average} partial effect on the predicted response, we instead plot the $N$ estimated conditional expectation curves: each reflects the predicted response as a function of covariate $x_S$, conditional on an observed $\x_C$.

Consider the observations $\braces{\parens{x_{Si},\,\x_{Ci}}}_{i=1}^N$, and the estimated response function $\hat{f}$. For each of the $N$ observed and fixed values of $\x_C$, a curve $\hat{f}^{(i)}_S$ is plotted against the observed values of $x_S$. Therefore, at each x-coordinate, $x_S$ is fixed and the $\x_C$ varies across $N$ observations.  Each curve defines the conditional relationship between $x_S$ and $\hat{f}$ at fixed values of $\x_C$. Thus, the ICE algorithm gives the user insight into the several variants of conditional relationships estimated by the black box.

The ICE algorithm is given in Algorithm \ref{algo:ice} in Appendix \ref{app:algorithms}. Note that the  PDP curve is the average of the $N$ ICE curves and can thus be viewed as a form of post-processing. Although in this paper we focus on the case where $|S|=1$, the pseudocode is general.  All plots in this paper are produced using the \texttt{R} package \texttt{ICEbox}, available on CRAN.

\begin{figure}[htp]
\centering
\includegraphics[width=2.5in]{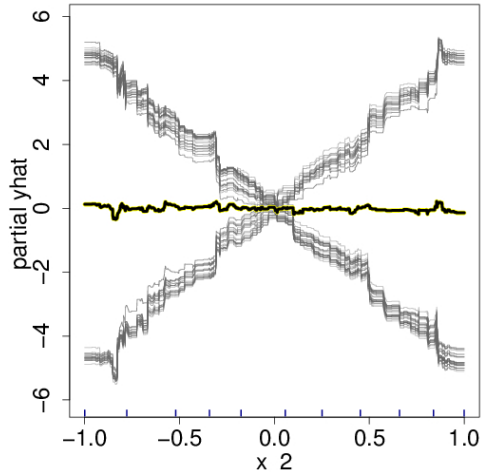}
\caption{\texttt{SGB} ICE plot for $X_2$ from 1000 realizations of the data generating process described by Equation \ref{eq:criss_cross_model0}.  We see that the \texttt{SGB}'s fitted values are either approximately linearly increasing or decreasing in $X_2$.}
\label{fig:criss_cross_x2_ice}
\end{figure}

Returning to the simulated data described by Equation \ref{eq:criss_cross_model0}, Figure \ref{fig:criss_cross_x2_ice} shows the ICE plot for the \texttt{SGB} when $S=X_2$.  In contrast to the PDP in Figure \ref{fig:criss_cross_x2_pdp}, the ICE plot makes it clear that the fitted values \emph{are} related to $X_2$.  Specifically, the \texttt{SGB}'s predicted values are approximately linearly increasing or decreasing in $X_2$ depending upon which region of $\mathcal{X}$ an observation is in.

\begin{figure}[htp]
\centering
\includegraphics[width=2.5in]{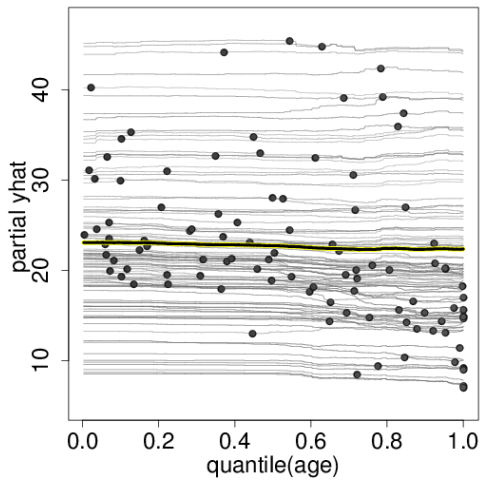}
\caption{\texttt{RF} ICE plot for BHD for predictor \texttt{age}. The highlighted thick line is the PDP. For each curve, the location of its observed \texttt{age} is marked by a point. For some observations, higher \texttt{age} is associated with a higher predicted values.  The upper set of tick marks on the horizontal axis indicate the observed deciles of \texttt{age}.}
\label{fig:bh_ice_example}
\end{figure}

Now consider the well known Boston Housing Data (BHD). The goal in this dataset is to predict a census tract's median home price using features of the census tract itself. It is important to note that the median home prices for the tracts are truncated at 50, and hence one may observe potential ceiling effects when analyzing the data.   We use Random Forests (\texttt{RF}) implemented in \texttt{R} \citep{Liaw2002} to fit $\hat{f}$. The ICE plot in Figure \ref{fig:bh_ice_example} examines the association between the average age of homes in a census tract and the corresponding median home value for that tract ($S=\texttt{age}$). The PDP is largely flat, perhaps displaying a slight decrease in predicted median home price as \texttt{age} increases.  The ICE plot shows those observations for which increasing \texttt{age} is actually associated with higher predicted values, thereby describing how individual behavior departs from the average behavior.

\subsection{The Centered ICE Plot}\label{subsec:centered_plot}

When the curves have a wide range of intercepts and are consequently ``stacked'' on each other, heterogeneity in the model can be difficult to discern. In Figure \ref{fig:bh_ice_example}, for example, the variation in effects between curves and cumulative effects are veiled.  In such cases the \qu{centered ICE} plot (the \qu{c-ICE}), which removes level effects, is useful.

c-ICE works as follows. Choose a location $x^*$ in the range of $x_S$ and join or \qu{pinch} all prediction lines at that point. We have found that choosing $x^*$ as the minimum or the maximum observed value results in the most  interpretable plots. For each curve $\hat{f}^{(i)}$ in the ICE plot,  the corresponding c-ICE curve is given by 

\bneqn \label{eq:c-ice}
\hat{f}_{\mathrm{cent}}^{(i)} = \hat{f}^{(i)} - \onevec \hat{f}(x^*,\x_{Ci}),
\eneqn

\noindent where the unadorned $\hat{f}$ denotes the fitted model and $\onevec$ is a vector of 1's of the appropriate dimension.  Hence the point $(x^*,\hat{f}(x^*,\x_{Ci}))$ acts as a ``base case'' for each curve.  If $x^*$ is the minimum value of $x_S$, for example, this ensures that all curves originate at 0, thus removing the differences in level due to the different $\x_{Ci}$'s. At the maximum $x_S$ value, each centered curve's level reflects the cumulative effect of $x_S$ on $\hat{f}$ relative to the base case.  The result is a plot that better isolates the combined effect of $x_S$ on $\hat{f}$, holding $\x_C$ fixed.

\begin{figure}[htp]
\centering
\includegraphics[width=2.5in]{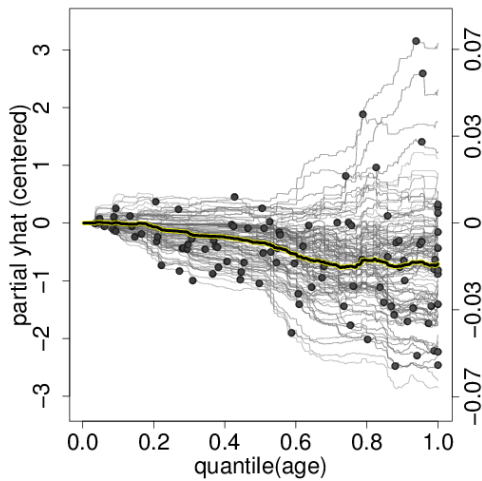}
\caption{c-ICE plot for \texttt{age} with $x^*$ set to the  minimum value of \texttt{age}.  The right vertical axis displays changes in $\hat{f}$ over the baseline as a fraction of $y$'s observed range. In this example, interactions between \texttt{age} and other predictors create cumulative differences in fitted values of up to about 14\% of the range of $y$.}
\label{fig:bh_c-ice_example}
\end{figure}

Figure \ref{fig:bh_c-ice_example} shows a c-ICE plot for the predictor \texttt{age} of the BHD for the same \texttt{RF} model as examined previously.  From the c-ICE plot we can now see clearly that the cumulative effect of \texttt{age} on predicted median value increases for some cases, and decreases for others. Such divergences of the centered curves suggest the existence of interactions between $x_S$ and $\x_C$ in the model. Also, the magnitude of the effect, as a fraction of the range of $y$, can be seen in the vertical axis displayed on the right of the graph.

\subsection{The Derivative ICE Plot}\label{subsec:derivative_plot}

To further explore the presence of interaction effects, we develop plots of the partial derivative of $\hat{f}$ with respect to $x_S$. To illustrate, consider the scenario in which $x_S$ does not interact with the other predictors in the fitted model.  This implies $\hat{f}$ can be written as

\bneqn\label{eq:additive_model}
\hat{f}(\x) = \hat{f}(x_S, \x_{C}) = g(x_S) + h(\x_{C}), \quad \text{so that} \quad \frac{\partial \hat{f}(\x)}{\partial{x_S}} = g'(x_S),
\eneqn

\noindent meaning the relationship between $x_S$ and $\hat{f}$ does not depend on $\x_{C}$.   Thus the ICE plot for $x_S$ would display a set of $N$ curves that share a single common shape but differ by level shifts according to the values of $\x_{C}$.

As it can be difficult to visually assess derivatives from ICE plots, it is useful to plot an estimate of the partial derivative directly.  The details of this procedure are given in Algorithm \ref{algo:nppb} in Appendix \ref{app:algorithms}.  We call this a \qu{derivative ICE} plot, or \qu{d-ICE.}  When no interactions are present in the fitted model, all curves in the d-ICE plot are equivalent, and the plot shows a single line.  When interactions do exist, the derivative lines will be heterogeneous.

\begin{figure}[htp]
\centering
\includegraphics[width=2.7in]{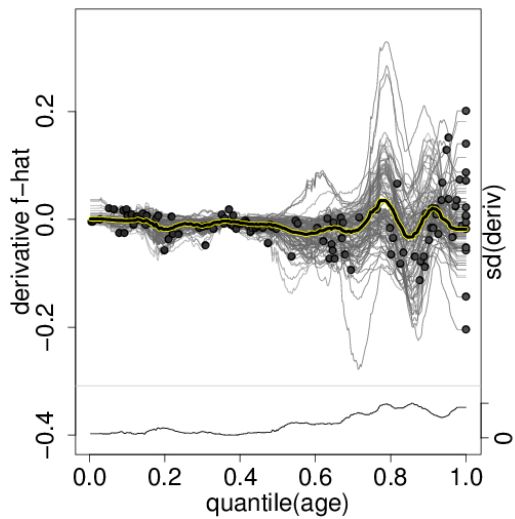}
\caption{d-ICE plot for \texttt{age} in the BHD. The left vertical axis' scale  gives the partial derivative of the fitted model. Below the d-ICE plot we plot the standard deviation of the derivative estimates at each value of \texttt{age}.  The scale for this standard deviation plot is on the bottom of the right vertical axis.}
\label{fig:bh_d-ice_example}
\end{figure}

As an example, consider the d-ICE plot for the \texttt{RF} model in Figure \ref{fig:bh_d-ice_example}.  The plot suggests that when \texttt{age} is below approximately 60, $g' \approx 0$ for all observed values of $\x_C$.  In contrast, when \texttt{age} is above 60 there are observations for which $g' > 0$ and others for which $g' < 0$, suggesting an interaction between \texttt{age} and the other predictors. Also, the standard deviation of the partial derivatives at each point, plotted in the lower panel, serves as a useful summary to highlight regions of heterogeneity in the estimated derivatives (i.e., potential evidence of interactions in the fitted model).

\subsection{Visualizing a Second Feature}\label{subsec:second_feature}

Color allows overloading of ICE, c-ICE and d-ICE plots with information regarding a second predictor of interest $x_k$. Specifically, one can assess how the second predictor influences the relationship between $x_S$ and $\hat{f}$. If $x_k$ is categorical, we assign colors to its levels and plot each prediction line $\hat{f}^{(i)}$ in the color of $x_{ik}$'s level. If $x_k$ is continuous, we vary the color shade from light (low $x_k$) to dark (high $x_k$).

\begin{figure}[htp]
\centering
\includegraphics[width=2.5in]{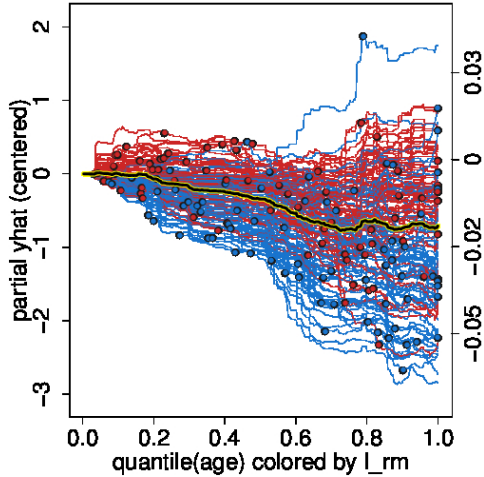}
\caption{The c-ICE plot for \texttt{age} of Figure \ref{fig:bh_c-ice_example} in the BHD. Red lines correspond to observations with \texttt{rm} greater than the median \texttt{rm} and blue lines correspond to those with fewer.}
\label{fig:bh_c-ice_colored_example}
\end{figure}

We replot the c-ICE from Figure \ref{fig:bh_c-ice_example} with lines colored by a newly constructed predictor, $x=1(\texttt{rm} > \mathrm{median}(\texttt{rm}))$. Lines are colored red if the average number of rooms in a census tract is greater than the median number of rooms across all all census tracts and are colored blue otherwise. Figure \ref{fig:bh_c-ice_colored_example} suggests that for census tracts with a larger number of average rooms, predicted median home price value is positively associated with \texttt{age} and for census tracts with a lesser number of average rooms, the association is negative. 

\section{Simulations}\label{sec:simulations}

Each of the following examples is designed to emphasize a particular model characteristic that the ICE toolbox can detect.  To more clearly demonstrate given scenarios with minimal interference from issues that one typically encounters in actual data, such as noise and model misspecification, the examples are purposely stylized. 

\subsection{Additivity Assessment}\label{subsec:sim_additivity}

We begin by showing that ICE plots can be used as a diagnostic in evaluating the extent to which a fitted model $\hat{f}$ fits an additive model.

Consider again the prediction task in which $\hat{f}(\x) = g(x_S) + h(\x_{C})$.  For arbitrary vectors $\x_{Ci}$ and $\x_{Cj}$, $\hat{f}(x_S, \x_{Ci}) - \hat{f}(x_S, \x_{Cj}) = h(\x_{Ci}) - h(\x_{Cj})$ for all values of $x_S$.  The term $h(\x_{Ci}) - h(\x_{Cj})$ represents the shift in level due to the difference between $\x_{Ci}$ and $\x_{Cj}$ and is independent of the value of $x_S$.  Thus the ICE plot for $x_S$ will display a set of $N$ curves that share a common shape but differ by level shifts according to the unique values of $\x_{C}$.

As an illustration, consider the following additive data generating model

\beqn
Y = X_{1}^2 + X_{2} + \errorrv, \quad X_{1}, X_{2} \iid \uniform{-1}{1}, \quad \errorrv \iid \normnot{0}{1}.
\eeqn

We simulate 1000 independent $(\X_i, Y_i)$ pairs according to the above and fit a generalized additive model \citep[\texttt{GAM},][]{Hastie1986} via the \texttt{R} package \texttt{gam} \citep{Hastie2013}.  As we have specified it, the \texttt{GAM} assumes 

\beqn
f(\X)=f_1(X_1)+f_2(X_2)+f_3(X_1 X_2)
\eeqn

\noindent where $f_1$, $f_2$ and $f_3$ are unknown functions estimated internally by the procedure using smoothing splines. Because $f_3$ appears in the model specification but not in the data generating process, any interaction effects that \texttt{GAM} fits are spurious.\footnote{If we were to eliminate $f_3$ from the \texttt{GAM} then we would know a priori that $\hat{f}$ would not display interaction effects.}  Here, ICE plots inform us of the degree to which interactions were fit. Were there no interaction in $\hat{f}$ between $X_1$ and $X_2$, the ICE plots for $X_1$ would display a set of curves equivalent in shape but differing in level.  

\begin{figure}[htp]
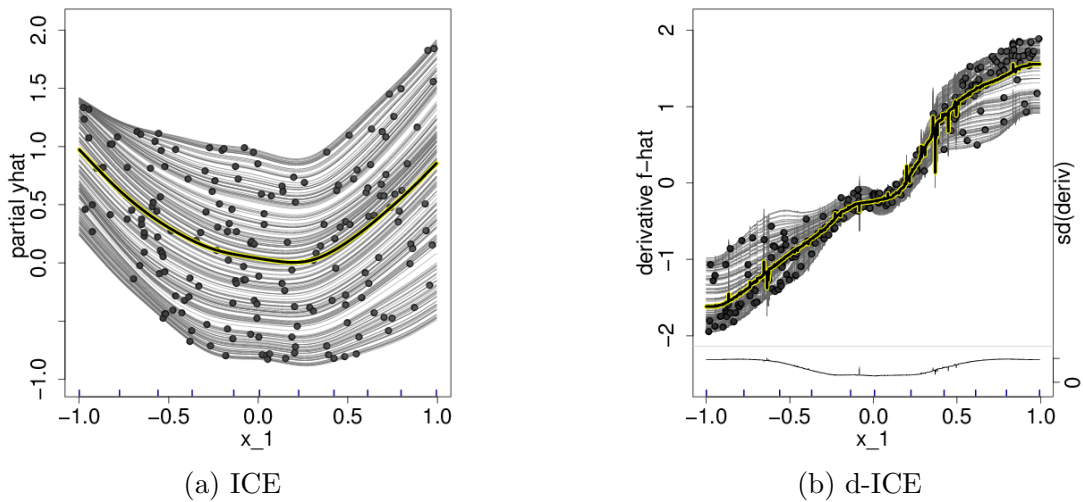

\centering
\begin{subfigure}[b]{0.48\textwidth}
                \centering
                \includegraphics[width=2.35in]{no_inter_example.\iftoggle{usejpgs}{jpg}{pdf}}
                \caption{ICE}
                \label{fig:no_inter_ice}
        \end{subfigure}~~
\begin{subfigure}[b]{0.48\textwidth}
                \centering
                \includegraphics[width=2.35in]{no_inter_example_d-ice.\iftoggle{usejpgs}{jpg}{pdf}}
                \caption{d-ICE}
                \label{fig:no_inter_d-ice}
        \end{subfigure}
\caption{ICE and d-ICE plots for $S=X_1$ when $\hat{f}$ is a \texttt{GAM} with possible interaction effects between $X_1$ and $X_2$.  So as to keep the plot uncluttered we plot only a fraction of all 1000 curves.  In the ICE plots the dots indicate the actual location of $X_1$ for each curve.}
\label{fig:no_inter_example}
\end{figure}

Figure \ref{fig:no_inter_ice} displays the ICE plots for $X_1$ and indicates that this is indeed the case: all curves display a similar parabolic relationship between $\hat{f}$ and $X_1$, shifted by a constant, and independent of the value of $X_2$. Accordingly, the associated d-ICE plot in Figure \ref{fig:no_inter_d-ice} displays little variation between curves. The ICE suite makes it apparent that $f_3$ (correctly) contributes relatively little to the \texttt{GAM} model fit. Note that additive structure cannot be observed from the PDP alone in this example (or any other).

\subsection{Finding interactions and regions of interactions}\label{subsec:sim_interactions}

As noted in \citet{Friedmana}, the PDP is most instructive when there are no interactions between $x_S$ and the other features. In the presence of interaction effects, the averaging procedure in the PDP can obscure any heterogeneity in $\hat{f}$. Let us return to the simple interaction model

\begin{figure}[htp]
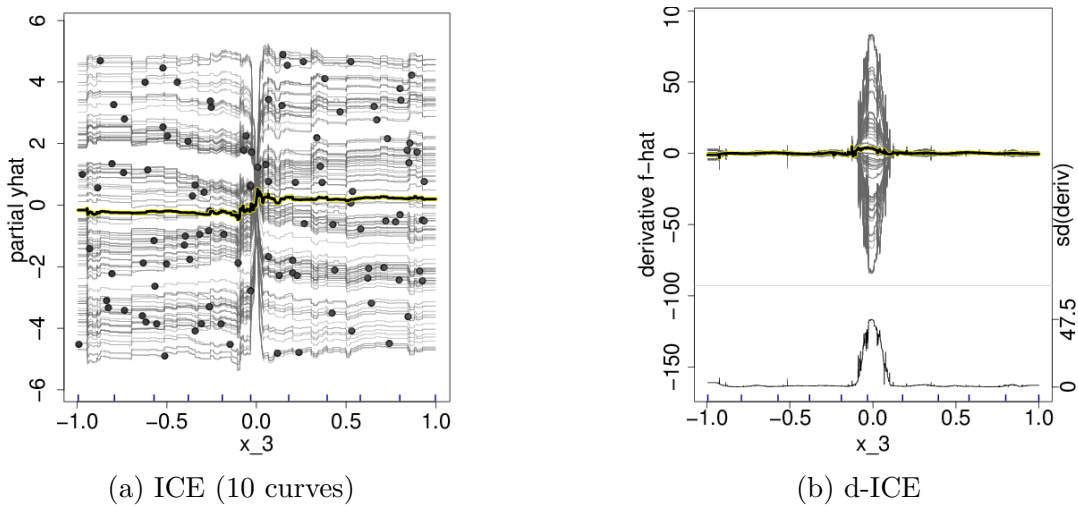

\centering
\begin{subfigure}[b]{0.48\textwidth}
                \centering
                \includegraphics[width=2.35in]{criss_cross_sim_gbm.\iftoggle{usejpgs}{jpg}{pdf}}
                \caption{ICE (10 curves)}
                \label{fig:criss_cross_sim_amdps_gbm}
        \end{subfigure}~~
\begin{subfigure}[b]{0.48\textwidth}
                \centering
                \includegraphics[width=2.35in]{criss_cross_sim_gbm_d.\iftoggle{usejpgs}{jpg}{pdf}}
                \caption{d-ICE}
                \label{fig:criss_cross_sim_amdps_gbm_d}
        \end{subfigure}
\caption{ICE plots for an \texttt{SGB} fit to the simple interaction model of Equation \ref{eq:criss_cross_model}.}
\label{fig:criss_cross_sim_amdps}
\end{figure}

\bneqn\label{eq:criss_cross_model}
&& Y = 0.2 X_{1} - 5X_{2}  + 10X_{2} \indic{X_{3} \geq 0} + \errorrv, \\ \nonumber
&& \errorrv \iid \normnot{0}{1}, \quad X_{1}, X_{2}, X_{3} \iid \uniform{-1}{1}
\eneqn

\noindent to examine the relationship between \texttt{SGB}'s $\hat{f}$ and $X_3$.  Figure \ref{fig:criss_cross_sim_amdps_gbm} displays an ICE plot for $X_3$. Similar to the PDP we saw in Section \ref{sec:introduction}, the plot suggests that averaged over $X_1$ and $X_2$, $\hat{f}$ is not associated with $X_3$. By following the non-parallel ICE curves, however, it is clear that $X_3$ modulates the fitted value through interactions with $X_1$ and $X_2$.

Where in the range of $X_3$ do these interactions occur? The d-ICE plot of Figure \ref{fig:criss_cross_sim_amdps_gbm_d} shows that interactions are in a neighborhood around $X_3 \approx 0$. This is expected; in the model given by Equation \ref{eq:criss_cross_model}, being above or below $X_3 = 0$ changes the response level. The plot suggests that the fitted model's interactions are concentrated in $X_3 \in \bracks{-0.025,\,0.025}$ which we call the \qu{region of interaction} (ROI). 

Generally, ROIs are identified by noting where the derivative lines are variable. In our example, the lines have highly variable derivatives (both positive and negative) in $\bracks{-0.025,\,0.025}$. The more heterogeneity in these derivative lines, the larger the effect of the interaction between $x_S$ and $\x_C$  on the model fit. ROIs can be seen most easily by plotting the standard deviation of the derivative lines at each $x_S$ value. In this example, the standard deviation function is plotted in the bottom pane of Figure \ref{fig:criss_cross_sim_amdps_gbm_d} and demonstrates that fitted interactions peak at $X_3 \approx 0$.

\subsection{Extrapolation Detection}\label{subsec:sim_extrap}

As the number of predictors $p$ increases, the sample vectors $\x_1,\ldots \x_N$ are increasingly sparse in the feature space $\mathcal{X}$.  A consequence of this curse of dimensionality is that for many $\x \in \mathcal{X}$, $\hat{f}(\x)$ represents an extrapolation rather than an interpolation (see \citealp{Hastie2009} for a more complete discussion).

Extrapolation may be of particular concern when using a black-box algorithm to forecast $\x_{\text{new}}$. Not only may $\hat{f}(\x_{\text{new}})$ be an extrapolation of the $(\x,y)$ relationship observed in the training data, but the black-box nature of $\hat{f}$ precludes us from gaining any insight into what the extrapolation might look like.  Fortunately, ICE plots can cast light into these extrapolations.

Recall that each curve in the ICE plot includes the fitted value $\hat{f}(x_{Si}, \x_{Ci})$ where $x_{Si}$ is actually observed in the training data for the $i$th observation. The other points on this curve represent extrapolations in $\mathcal{X}$.  Marking each curve in the ICE plot at the observed point helps us assess the presence and nature of $\hat{f}$'s hypothesized extrapolations in $\mathcal{X}$.

Consider the following model:

\bneqn \label{eq:extrap}
&& Y = 10X_1^2 + \indic{X_{2} \geq 0} + \errorrv, \\ 
&& \errorrv \iid \normnot{0}{.1^2}, \quad \twovec{X_1}{X_2} \sim \begin{cases}
\uniform{-1}{0}, ~~\uniform{-1}{0}  & \withprob \third \\
\uniform{0}{1}, ~~\uniform{-1}{0}   & \withprob \third \\ \nonumber
\uniform{-1}{0}, ~~\uniform{0}{1}   & \withprob \third \\ \nonumber
\uniform{0}{1}, ~~\uniform{0}{1}   & \withprob 0.
\end{cases}
\eneqn

\begin{figure}[htp]
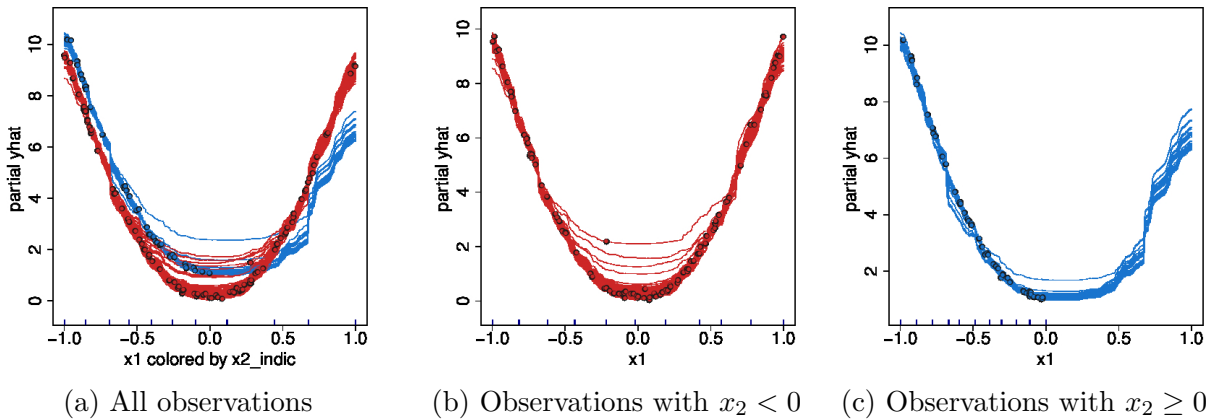

\centering
\begin{subfigure}[b]{0.32\textwidth}
                \centering
                \includegraphics[width=1.9in]{extrap_sim_full_red.\iftoggle{usejpgs}{jpg}{pdf}}
                \caption{All observations}
                \label{fig:extrapsim a}
        \end{subfigure}~~
\begin{subfigure}[b]{0.32\textwidth}
                \centering
                \includegraphics[width=1.9in]{extrap_sim_noextrap_red.\iftoggle{usejpgs}{jpg}{pdf}}
                \caption{Observations with $x_2<0$}
                \label{fig:extrapsim b}
        \end{subfigure}
\begin{subfigure}[b]{0.32\textwidth}
                \centering
                \includegraphics[width=1.9in]{extrap_sim_extrap_blue.\iftoggle{usejpgs}{jpg}{pdf}}
                \caption{Observations with $x_2 \geq 0$}
                \label{fig:extrapsim c}
        \end{subfigure}        
\caption{ICE plots for $S=x_1$ of a \texttt{RF} model fit to Equation \ref{eq:extrap}.  The left plot shows the ICE plot for the entire dataset where $x_2 < 0$ is colored red and $x_2 \geq 0$ in blue.  The middle plot shows only the red curves and the right only the blue. Recall that there is no training data in the quadrant $[0, 1] \times [0, 1]$, and so Figure \ref{fig:extrapsim c} contains no points for observed values when $x_1 > 0$ (when both $x_1$ and $x_2$ are positive).  Nevertheless, from Figure \ref{fig:extrapsim c}'s ICE curves it is apparent that the fitted values are increasing in $x_1$ for values above $0$.  Here, the ICE plot elucidates the existence and nature of the \texttt{RF}'s extrapolation outside the observed $\mathcal{X}$-space.}
\label{fig:extrapsim}
\end{figure}

Notice $\prob{X_1>0,\,X_2>0}=0$, leaving the quadrant $[0, 1] \times [0, 1]$ empty.  We simulate 1000 observations and fit a \texttt{RF} model to the data. The ICE plot for $x_1$ is displayed in Figure \ref{fig:extrapsim a} with the points corresponding to the 1000 observed $(x_1,\,x_2)$ values marked by dots.  We highlight observations with $x_2 < 0$ in red and those with $x_2 \geq 0$ in blue.  The two subsets are plotted separately in Figures \ref{fig:extrapsim b} and \ref{fig:extrapsim c}.

The absence on the blue curves of points where both $x_1,x_2 > 0$ confirms that the probability of $X_1 > 0$ and $X_2 > 0$ equals zero. From Figure \ref{fig:extrapsim c}, we see that in this region of $\mathcal{X}$, $\hat{f}$ increases roughly in proportion with $x_1^2$ even though no data exists.  Ostensibly the \texttt{RF} model has extrapolated the polynomial relationship from the observed $\mathcal{X}$-space to where both $x_1 > 0$ and $x_2 > 0$.

Whether it is desirable for $\hat{f}$ to display such behavior in unknown regions of $\mathcal{X}$ is dependent on the character of the extrapolations in conjunction with the application at hand.  Moreover, different algorithms will likely give different extrapolations.  Examining the ICE plots can reveal the nature of these extrapolations and guide the user to a suitable choice.

\section{Real Data}\label{sec:real_data}

We now demonstrate the ICE toolbox on three real data examples. We emphasize features of $\hat{f}$ that might otherwise have been overlooked.

\subsection{Depression Clinical Trial}\label{subsec:depression}

The first dataset comes from a depression clinical trial \citep{DeRubeis2013}. The response variable is the Hamilton Depression Rating Scale (a common composite score of symptoms of depression where lower scores correspond to being less depressed) after 15 weeks of treatment. The treatments are placebo, cognitive therapy (a type of one-on-one counseling), and paroxetine (an anti-depressant medication). The study also collected 37 covariates which are demographic (e.g. age, gender, income) or related to the medical history of the subject (e.g. prior medications and whether the subject was previously treated). For this illustration, we drop the placebo subjects to focus on the 156 subjects who received either of the two active treatments.

The goal of the analysis in \citet{DeRubeis2013} is to understand how different subjects respond to different treatments, conditional on their \textit{personal} covariates. The difference between the two active treatments, assuming the classic linear (and additive) model for treatment, was found to be statistically insignificant.  If the clinician believes that the treatment effect is heterogeneous and the relationship between the covariates and response is complex, then flexible nonparametric models could be an attractive exploratory tool.

Using the ICE toolbox, one can visualize the impact of the treatment variable on an $\hat{f}$ given by a black box algorithm. Note that extrapolations in the treatment indicator (i.e. predicting at 0 for an observed 1 or vice versa) correspond to counterfactuals in a clinical setting, allowing the researcher to see how the same patient might have responded to a different treatment.

We first modeled the response as a function of the 37 covariates as well as treatment to obtain the best fit of the functional relationship using the black-box algorithm \texttt{BART} (implemented by \citealp{Kapelner2013}) and obtained an in-sample $R^2 \approx 0.40$.

\begin{figure}[htp]
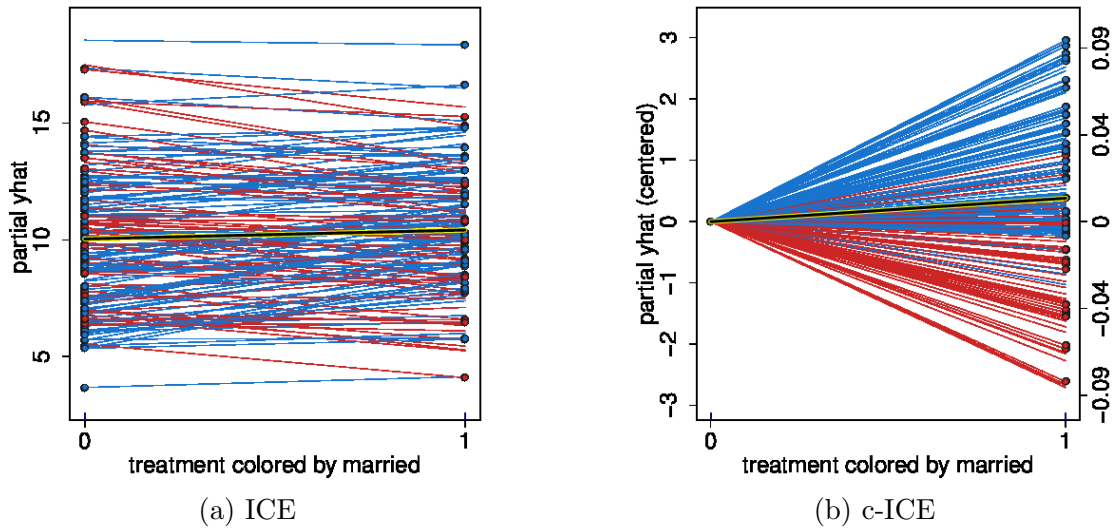

\centering
\begin{subfigure}[b]{0.48\textwidth}
                \centering
                \includegraphics[width=2.5in]{cog_therapy_married_red.\iftoggle{usejpgs}{jpg}{pdf}}
                \caption{ICE}
                \label{fig:depression}
        \end{subfigure}~~
\begin{subfigure}[b]{0.48\textwidth}
                \centering
                \includegraphics[width=2.5in]{cog_therapy_married_red_c.\iftoggle{usejpgs}{jpg}{pdf}}
                \caption{c-ICE}
                \label{fig:depression_c}
        \end{subfigure}
\caption{ICE plots of a \texttt{BART} model for the effect of treatment on depression score after 15 weeks.  Married subjects are colored in blue and unmarried subjects are colored in red.}
\label{fig:depression_ices}
\end{figure}

Figure \ref{fig:depression} displays an ICE plot of the binary treatment variable, with cognitive therapy coded as ``0'' and paroxetine coded as ``1'', colored by marital status (blue if married and red if unmarried). The plot shows a flat PDP, demonstrating no relationship between the predicted response and treatment when averaging over the effects of other covariates. However, the crossing of ICE curves indicates the presence of interactions in $\hat{f}$, which is confirmed by the c-ICE plot in Figure \ref{fig:depression_c}. After centering, it becomes clear that the flat PDP obscures a complex relationship: the model predicts between -3 and +3 points on the Hamilton scale, which is a highly clinically significant range (and almost 20\% of the observed response's range).  Further, we can see that \texttt{BART} fits an interaction between treatment and marital status: married subjects are generally predicted to do better on cognitive therapy and unmarried subjects are predicted to do better with paroxetine.  

\subsection{White Wine}\label{subsec:white_wine}

The second data set concerns 5,000 white wines produced in the \textit{vinto verde} region of Portugal obtained from the UCI repository \citep{Bache2013}. The response variable is a wine quality metric, taken to be the median preference score of three blind tasters on a scale of 1-10, treated as continuous. The 11 covariates are physicochemical metrics that are commonly collected for wine quality control such as citric acid content, sulphates, etc. The model is fit with a neural network (\texttt{NN}) using the R package \texttt{nnet} \citep{Venables2002}. We fit a \texttt{NN} with 3 hidden units and a small parameter value for weight decay\footnote{Note that \texttt{NN} models are highly sensitive to the number of hidden units and weight decay parameter.  We therefore offer the following results as merely representative of the type of plots which \texttt{NN} models can generate.} and achieved an in-sample $R^2$ of approximately 0.37.

\begin{figure}[htp]
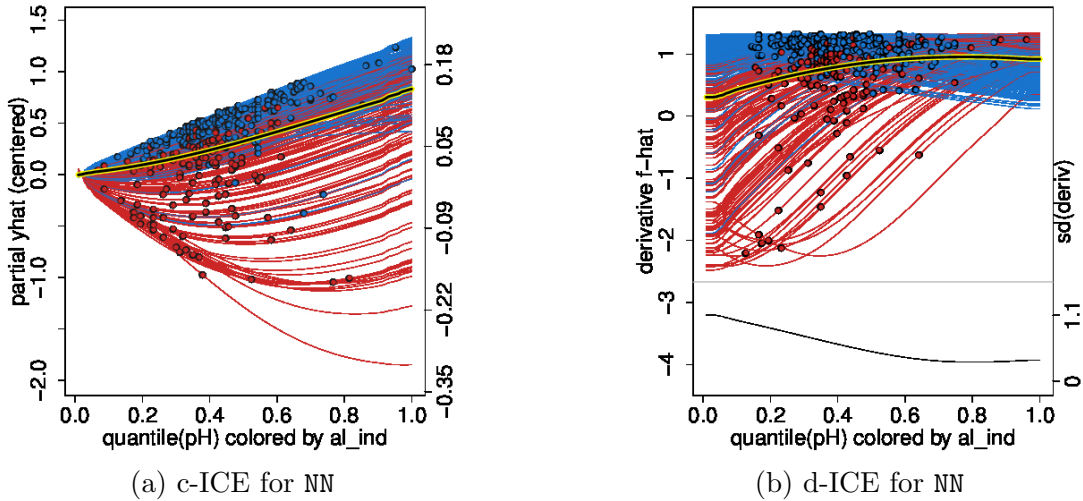

\centering
\begin{subfigure}[b]{0.48\textwidth}
                \centering
                \includegraphics[width=2.35in]{white_wine_ph_by_median_alcohol_c_nnets.\iftoggle{usejpgs}{jpg}{pdf}}
                \caption{c-ICE for \texttt{NN}}
                \label{fig:white_wine_c_nn}
        \end{subfigure}~~
\begin{subfigure}[b]{0.48\textwidth}
                \centering
                \includegraphics[width=2.35in]{white_wine_ph_by_median_alcohol_d_nnets.\iftoggle{usejpgs}{jpg}{pdf}}
                \caption{d-ICE for \texttt{NN}}
                \label{fig:white_wine_d_nn}
        \end{subfigure}
\caption{ICE plots of \texttt{NN} model for wine ratings versus \texttt{pH} of white wine colored by whether the alcohol content is high (blue) or low (red). To prevent cluttering, only a fraction of the 5,000 observations are plotted.}
\label{fig:white_wine}
\end{figure}

We find the covariate \texttt{pH} to be the most illustrative. The c-ICE plot is displayed in Figure \ref{fig:white_wine_c_nn}. Wines with high alcohol content are colored blue and wines with low alcohol content are colored red. Note that the PDP shows a linear trend, indicating that on average, higher \texttt{pH} is associated with higher fitted preference scores. While this is the general trend for wines with higher alcohol content, the ICE plots reveal that interaction effects are present in $\hat{f}$. For many white wines with low alcohol content, the illustration suggests a nonlinear and cumulatively \textit{negative} association. For these wines, the predicted preference score is actually negatively associated with \texttt{pH} for low values of \texttt{pH} and then begins to increase --- a severe departure from what the PDP suggests.  However, the area of increase contains no data points, signifying that the increase is merely an extrapolation likely driven by the positive trend of the high alcohol wines. Overall, the ICE plots indicate that for more alcoholic wines, the predicted score is increasing in \texttt{pH} while the opposite is true for wines with low alcohol content. Also, the difference in cumulative effect is meaningful; when varied from the minimum to maximum values of \texttt{pH}, white wine scores vary by roughly 40\% of the range of the response variable.

Examining the derivative plot of Figure \ref{fig:white_wine_d_nn} confirms the observations made above. The \texttt{NN} model suggests interactions exist for lower values of \texttt{pH} in particular. Wines with high alcohol content have mostly positive derivatives while those with low alcohol content have mostly negative derivatives. As \texttt{pH} increases, the standard deviation of the derivatives decreases, suggesting that interactions are less prevalent at higher levels of \texttt{pH}.  

\subsection{Diabetes Classification in Pima Indians}\label{subsec:pima}

The last dataset consists of 332 Pima Indians \citep{Smith1988} obtained from the \texttt{R} library \texttt{MASS}.  Of the 332 subjects, 109 were diagnosed with diabetes, the binary response variable which was fit using seven predictors (with body metrics such as blood pressure, glucose concentration, etc.). We model the data using a \texttt{RF} and achieve an out-of-bag misclassification rate of 22\%.

\begin{figure}[htp]
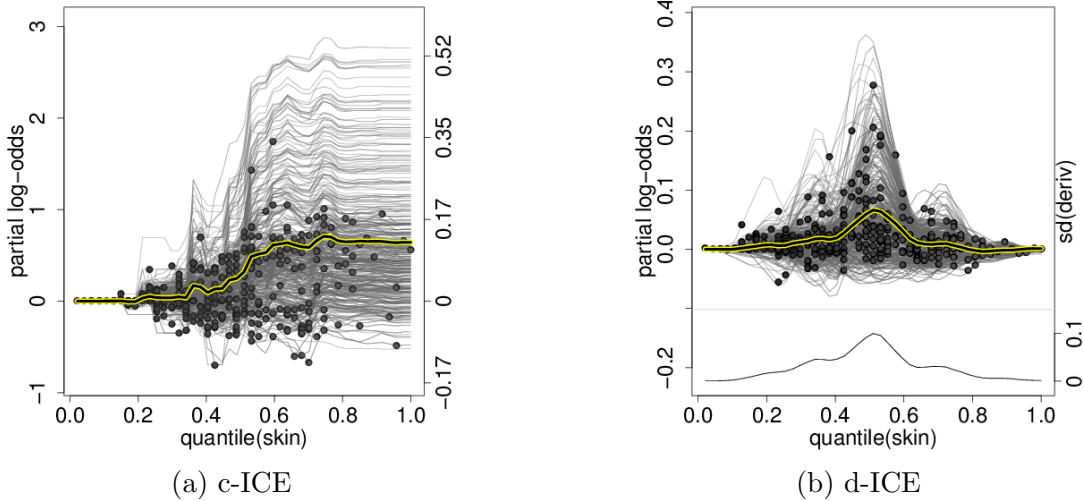

\centering
\begin{subfigure}[b]{0.48\textwidth}
                \centering
                \includegraphics[width=2.35in]{pima_skin_by_age_c.\iftoggle{usejpgs}{jpg}{pdf}}
                \caption{c-ICE}
                \label{fig:pima_skin_age_c}
        \end{subfigure}~~
\begin{subfigure}[b]{0.48\textwidth}
                \centering
                \includegraphics[width=2.35in]{pima_skin_by_age_d.\iftoggle{usejpgs}{jpg}{pdf}}
                \caption{d-ICE}
                \label{fig:pima_skin_age_d}
        \end{subfigure}
\caption{ICE plots of a \texttt{RF} model for estimated centered logit of the probability of contracting diabetes versus \texttt{skin} colored by subject \texttt{age}.}
\label{fig:pima_skin_age}
\end{figure}

Once again, ICE plots offer the practitioner a more comprehensive view of the output of the black box. For example, the covariate \texttt{skin} thickness about the triceps is plotted as a c-ICE in Figure \ref{fig:pima_skin_age_c}. The PDP clearly shows an increase in the predicted centered log odds of contracting diabetes. This is expected given that  \texttt{skin} is a proxy for obesity, a major risk factor for diabetes. However, the ICE plot illustrates a more elaborate model fit. Many subjects with high \texttt{skin} have a flat risk of diabetes according to $\hat{f}$; others with comparable thickness exhibit a much larger centered log-odds increase.\footnote{The curves at the top of the figure mainly correspond to younger people. Their estimated effect of high thickness is seen to be an extrapolation.}  Figure \ref{fig:pima_skin_age_d} shows that the \texttt{RF} model fits interactions across the range of \texttt{skin} with the largest heterogeneity in effect occurring when \texttt{skin} is slightly above 30.  This can be seen in the standard deviation of the derivative in the bottom pane of Figure \ref{fig:pima_skin_age_d}.


\section{A Visual Test for Additivity}\label{sec:testing}

Thus far we have used the ICE toolbox to explore the output of black box models. We have explored whether $\hat{f}$ has additive structure or if interactions exist, and also examined $\hat{f}$'s extrapolations in $\mathcal{X}$-space. To better visualize interactions, we plotted individual curves in colors according to the value of a second predictor $x_k$. We have \emph{not} asked whether these findings are reflective of phenomena in any underlying model.

When heterogeneity in ICE plots is observed, the researcher can adopt two mindsets.  When one considers $\hat{f}$ to be the fitted model used for subsequent predictions, the heterogeneity is of interest because it determines future fitted values.  This is the mindset we have considered thus far.  Separately, it might be interesting to ascertain whether interactions between $x_S$ and $\x_C$ exist in the data generating model, denoted $f$. This question exists for other discoveries made using ICE plots, but we focus here on interactions.

The problem of assessing the statistical validity of discoveries made by examining plots is addressed in \citet{Buja2009} and \citet{Wickham2010}.  The central idea in these papers is to insert the observed plot randomly into a lineup of null plots generated from data sampled under a null distribution. If the single real plot is correctly identified amongst 19 null plots, for example, then ``the discovery can be assigned a $p$-value of $0.05$" \citep{Buja2009}.  A benefit of this approach is that the procedure is valid despite the fact that we have not specified the form of the alternative distribution --- the simple instruction ``find the plot that appears different" is sufficient.  

\subsection{Procedure}
We adapt this framework to the specific problem of using ICE plots to evaluate additivity in a statistically rigorous manner.  For the exposition in this section, suppose that the response $y$ is continuous, the covariates $\x$ are fixed, and $y=f(\x) + \errorrv$.  Further assume $\expe{\errorrv}=0$ and  

\bneqn\label{eq:additive}
f(\x) = g(x_S) + h(\x_C),
\eneqn

\noindent meaning the true $\x$-conditional expectation of $y$ is additive in functions of $x_S$ and $\x_C$.  Let $F$ be the distribution of $\hat{f}$ when Equation \ref{eq:additive} holds and $f$ is additive.  We wish to test $H_0$: $\hat{f} \sim F$ versus $H_a$: $H_0$ is false.
  
Recall that ICE plots displaying non-parallel curves suggest that $\hat{f}$ is not additive in functions of $x_S$ and $\x_C$.  Thus if we can correctly identify a plot displaying such features amongst $K-1$ null plots generated under $F$, the discovery is valid at $\alpha = 1/K$.

We sample from $F$ by using backfitting \citep{Breiman1985} to generate  $\gstar$ and $\hstar$, estimates of $g$ and $h$, and then bootstrapping the residuals.  Both $\gstar$ and $\hstar$ can be obtained via any supervised learning procedures. The general procedure for $|S|=1$ proceeds is as follows.

\begin{enumerate}[1]
        \item Using backfitting, obtain $\gstar$ and $\hstar$. Then compute a vector of fitted values $\yhatstar = \gstar(x_S) + \hstar(\x_C)$ and a vector of residuals $\rstar := y - \yhatstar$.
        \item Let $\bv{r}_b$ be a random resampling of $\rstar$.  If heteroscedasticity is of concern, one can keep $\rstar$'s absolute values fixed and let $\bv{r}_b$ be a permutation of $\rstar$'s signs.  Define $\bv{y}_b:=\yhatstar+\bv{r}_b$. Note that $\cexpe{\bv{y}_b}{\x}$ is additive in $\gstar(x_S)$ and $\hstar(\x_{C})$.  
        \item Fit $\bv{y}_b$ to $\X$ using the same learning algorithm that generated the original ICE (c-ICE or d-ICE) plot to produce $\hat{f}_{b}$. This yields a potentially non-additive approximation to null data generated using an additive model.  
        \item Display an ICE (or c-ICE or d-ICE) plot for $\hat{f}_b$. Deviations from additivity observed in this plot must be due to sources other than interactions between $x_S$ and $\x_{C}$ in the underlying data.
        \item Repeat steps (2) - (4) $K - 1$ times, then randomly insert the true plot amongst these $K - 1$ null plots.
    \item If the viewer can correctly identify the true plot amongst all $K$ plots, the discovery is valid for level $\alpha = 1 / K$.  Note that the discovery is conditional on the procedures for generating $\gstar$ and $\hstar$.
\end{enumerate}

\subsection{Examples}\label{subsec:visualization_additivity_test}

\begin{figure}[htp]
\centering
\includegraphics[width=2.5in]{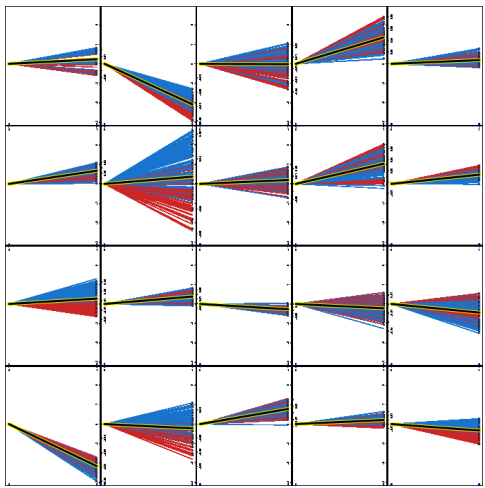} 
\caption{Additivity lineup test for the predictor \texttt{treatment} in the depression clinical trial dataset of Section \ref{subsec:depression}.}
\label{fig:depression_treatment_additivity_lineup_test}
\end{figure}

An application of this visual test where $g$ is taken to be the \qu{supersmoother} \citep{Friedman1984} and $h$ is a \texttt{BART} model is illustrated using the depression data of Section \ref{subsec:depression}. We sample $\bv{r}_b$ by permuting signs. The data analyst might be curious if the ICE plot is consistent with the treatment being additive in the model. We employ the additivity lineup test in Figure \ref{fig:depression_treatment_additivity_lineup_test} using 20 images. We reject the null hypothesis of additivity of the treatment effect at $\alpha = 1 / 20 = 0.05$ since the true plot (row 2, column 2) is clearly identifiable. This procedure can be a useful test in clinical settings when the treatment effect is commonly considered linear and additive and can alert the practitioner that interactions should be investigated. 

\begin{figure}[htp]
\centering
\includegraphics[width=2.5in]{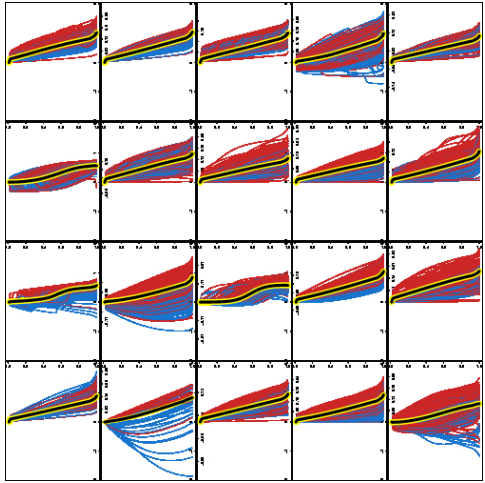}
\caption{Additivity lineup test for the predictor \texttt{pH} in the white wine dataset of Section \ref{subsec:white_wine}.}
\label{fig:wine_additivity_lineup_test}
\end{figure}

Another application of this visual test where $g$ is taken to be the supersmoother and $h$ is a \texttt{NN} model is illustrated using the wine data of Section \ref{subsec:white_wine}.  Here again we sample $\bv{r}_b$ by permuting signs. The data analyst may want to know if the fitted model is suggestive of interactions between \texttt{pH} and the remaining features in the underlying model. We employ the additivity lineup test in Figure \ref{fig:wine_additivity_lineup_test}, again using 20 images.

Looking closely one sees that the first and third plots in the last row have the largest range of cumulative effects and exhibit more curvature in individual curves than most of the other plots, making them the most extreme violations of the null.  Readers that singled out the first plot in the last row would have a valid discovery at $\alpha = .05$, but clearly the evidence of non-additivity is much weaker here than in the previous example.  Whereas Figure \ref{fig:depression_treatment_additivity_lineup_test} suggests the real plot is identifiable amongst more than 20 images, it would be easy to confuse Figure \ref{fig:wine_additivity_lineup_test}'s true plot with the one in row 4, column 3.  Hence there is only modest evidence that \texttt{pH}'s impact on $\hat{f}$ is different from what a \texttt{NN} might generate if there were no interactions between \texttt{pH} and the other predictors.

\section{Discussion}\label{sec:discussion}

We developed a suite of tools for visualizing the fitted values generated by an arbitrary supervised learning procedure. Our work extends the classical partial dependence plot (PDP), which has rightfully become a very popular visualization tool for black-box machine learning output. The partial functional relationship, however, often varies conditionally on the values of the other variables. The PDP offers the average of these relationships and thus individual conditional relationships are consequently masked, unseen by the researcher. These individual conditional relationships can now be visualized, giving researchers additional insight into how a given black box learning algorithm makes use of covariates to generate predictions.

The ICE plot, our primary innovation, plots an entire distribution of individual conditional expectation functions for a variable $x_S$. Through simulations and real data examples, we illustrated much of what can be learned about the estimated model $\hat{f}$ with the help of ICE. For instance, when the remaining features $\x_C$ do not influence the association between $x_S$ and $\hat{f}$, all ICE curves lie on top of another.  When $\hat{f}$ is additive in functions of $\x_C$ and $x_S$, the curves lie parallel to each other.  And when the partial effect of $x_S$ on $\hat{f}$ is influenced by $\x_{C}$, the curves will differ from each other in shape.  Additionally, by marking each curve at the $x_S$ value observed in the training data, one can better understand $\hat{f}$'s extrapolations. Sometimes these properties are more easily distinguished in the complementary \qu{centered ICE} (c-ICE) and \qu{derivative ICE} (d-ICE) plots. In sum, the suite of ICE plots provides a tool for visualizing an arbitrary fitted model's map between predictors and predicted values.

The ICE suite has a number of possible uses that were not explored in this work. While we illustrate ICE plots using the same data as was used to fit $\hat{f}$, out-of-sample ICE plots could also be valuable. For instance, ICE plots generated from random vectors in $\reals^p$ can be used to explore other parts of $\mathcal{X}$ space, an idea advocated by \citet{Plate2000}.  Further, for a single out-of-sample observation, plotting an ICE curve for each predictor can illustrate the sensitivity of the fitted value to changes in each predictor for this particular observation, which is the goal of the \qu{contribution plots} of \citet{Strumbelj2011}. Additionally, investigating ICE plots from $\hat{f}$'s produced by multiple statistical learning algorithms can help the researcher compare models. Exploring other functionality offered by the \texttt{ICEbox} package, such as the ability to cluster ICE curves, is similarly left for subsequent research.

The tools summarized thus far pertain to \emph{exploratory} analysis.  Many times the ICE toolbox provides evidence of interactions, but how does this evidence compare to what these plots would have looked like if no interactions existed?  Section \ref{sec:testing} proposed a \emph{testing} methodology. By generating additive models from a null distribution and introducing the actual ICE plot into the lineup, interaction effects can be distinguished from noise, providing a test at a known level of significance. Future work will extend the testing methodology to other null hypotheses of interest.

\section*{Supplementary Materials}\label{sec:replication} %
The procedures outlined in Section  \ref{sec:algorithm} are implemented in the \texttt{R} package \texttt{ICEbox} available on \texttt{CRAN}.  Simulated results, tables, and figures specific to this paper can be replicated via the script included in the supplementary materials.
The depression data of Section \ref{subsec:depression} cannot be released due to privacy concerns.

\section*{Acknowledgements}

We thank Richard Berk for insightful comments on multiple drafts and suggesting color overloading. We thank Andreas Buja for helping conceive the testing methodology. We thank Abba Krieger for his helpful suggestions. We also wish to thank Zachary Cohen for the depression data of Section \ref{subsec:depression} and helpful comments.  Alex Goldstein acknowledges support from the Simons Foundation Autism Research Initiative. Adam Kapelner acknowledges support from the National Science Foundation's Graduate Research Fellowship Program.

\bibliographystyle{apalike}\bibliography{ice_paper}

\appendix

\section{Algorithms}\label{app:algorithms}

\begin{algorithm}[htp]
\small
\caption{\small ICE algorithm: Given $\X$, the $N \times p$ feature matrix, $\hat{f}$, 
the fitted model, $S \subset \braces{1, \ldots, p}$, the subset of predictors for which to 
compute partial dependence, return $\hat{f}^{(1)}_S, \ldots, \hat{f}^{(N)}_S$, the estimated partial dependence curves for constant values of $\x_C$.}
\begin{algorithmic}[1]
\Function{ICE}{$\X$, $\hat{f}$, $S$}
        \For{$i \gets 1 \ldots  N$}
                \State $\hat{f}^{(i)}_S \gets \bv{0}_{N \times 1}$
                \State $\x_C \gets \X[i, C]$ \Comment{fix $\x_C$ at the $i$th observation's $C$ columns}
                \For{$\ell \gets 1 \ldots N$}                
                        \State $\x_S \gets \X[\ell, S]$ \Comment{vary $\x_S$}
                        \State $\hat{f}^{(i)}_{S \ell} \gets \hat{f}(\bracks{\x_S,~\x_C})$
                        \Comment{the $i$th curve's $\ell$th coordinate}
                \EndFor
        \EndFor
        \State \Return $[\hat{f}^{(1)}_{S}, \ldots, \hat{f}^{(N)}_{S}]$
\EndFunction
\end{algorithmic}
\label{algo:ice}
\end{algorithm}

\begin{algorithm}[htp]
\small
\caption{\small d-ICE algorithm: Given $\X$, the $N \times p$ feature matrix; $\hat{f}^{(1)}_S, \ldots, \hat{f}^{(N)}_S$, the estimated partial dependence functions for subset $S$ in the ICE plot; $D$, a function that computes the numerical derivative; returns $d\hat{f}^{(1)}_S, \ldots, d\hat{f}^{(N)}_S$, the derivatives of the estimated partial dependence.  In our implementation $D$ first smooths the ICE plot using the ``supersmoother'' and subsequently estimates the derivative from the smoothed ICE plot.}
\begin{algorithmic}[1]
\Function{d-ICE}{$\X, \hat{f}^{(1)}_S, \ldots, \hat{f}^{(N)}_S, D$}
        \For{$i \gets 1 \ldots N$}
                \State $d\hat{f}^{(i)}_S \gets \bv{0}_{N \times 1}$
                \State $\x_C \gets \X[i, C]$ \Comment{row of the $i$th observation, columns corresponding to $C$}
                \For{$\ell \gets 1 \ldots N$}
                        \State $\x_S \gets \X[\ell, S]$
                        \State $d\hat{f}^{(i)}_{S\ell} \gets D\bracks{\hat{f}^{(i)}(\x_S, \x_C)}$ \Comment{numerical partial derivative at $\hat{f}^{(i)}(\x_S, \x_C)$ w.r.t.  $\x_S$}
                \EndFor
        \EndFor
        \State \Return $[d\hat{f}^{(1)}_{S}, \ldots, d\hat{f}^{(N)}_{S}]$
\EndFunction
\end{algorithmic}
\label{algo:nppb}
\end{algorithm}

\end{document}

%% file: preamble.tex
\usepackage[auth-sc,affil-sl]{authblk}
\usepackage{graphicx}
\usepackage{color}
\usepackage{amsmath}
\usepackage{dsfont}
\usepackage[round]{natbib}
\usepackage{amssymb}  
\usepackage{abstract}
\usepackage{cancel}
\usepackage[margin=1in]{geometry}  
\usepackage{enumerate}
\usepackage{listings}
\usepackage{array}
\usepackage{algorithm}
\usepackage{float}
\usepackage{multirow}
\usepackage{algorithm}
\usepackage{algorithmicx}
\usepackage{algpseudocode}
\usepackage{subcaption}
\usepackage{etoolbox}

\newcommand{\qu}[1]{``#1''}

\newcounter{probnum}
\allowdisplaybreaks

\definecolor{gray}{rgb}{0.7,0.7,0.7}

\definecolor{black}{rgb}{0,0,0}
\definecolor{white}{rgb}{1,1,1}
\definecolor{blue}{rgb}{0,0,0.7}

\definecolor{green}{rgb}{0.133,0.545,0.133}

\definecolor{yellow}{rgb}{1,0.549,0}

\definecolor{red}{rgb}{1,0.133,0.133}

\definecolor{purple}{rgb}{0.58,0,0.827}

\definecolor{brown}{rgb}{0.55,0.27,0.07}

\definecolor{backgcode}{rgb}{0.97,0.97,0.8}
\definecolor{Brown}{cmyk}{0,0.81,1,0.60}
\definecolor{OliveGreen}{cmyk}{0.64,0,0.95,0.40}
\definecolor{CadetBlue}{cmyk}{0.62,0.57,0.23,0}

\newcommand{\bv}[1]{\boldsymbol{#1}}

\newcommand{\iid}{~{\buildrel iid \over \sim}~}

\newcommand{\third}{\frac{1}{3}}

\newcommand{\X}{\bv{X}}

\newcommand{\x}{\bv{x}}
\newcommand{\onevec}{\bv{1}}

\newcommand{\y}{\bv{y}}

\newcommand{\yhat}{\hat{\y}}

\newcommand{\twovec}[2]{\bracks{\begin{array}{c} #1 \\ #2 \end{array}}}

\newcommand{\reals}{\mathbb{R}}

\newcommand{\beqn}{\vspace{-0.25cm}\begin{eqnarray*}}
\newcommand{\eeqn}{\end{eqnarray*}}
\newcommand{\bneqn}{\vspace{-0.25cm}\begin{eqnarray}}
\newcommand{\eneqn}{\end{eqnarray}}

\newcommand{\parens}[1]{\left(#1\right)}

\newcommand{\prob}[1]{\mathbb{P}\parens{#1}}

\newcommand{\bracks}[1]{\left[#1\right]}
\newcommand{\braces}[1]{\left\{#1\right\}}

\newcommand{\expe}[1]{\mathbb{E}\bracks{#1}}
\newcommand{\cexpe}[2]{\expe{#1 ~ | ~ #2}}

\newcommand{\expesub}[2]{\mathbb{E}_{#1}\bracks{#2}}
\newcommand{\indic}[1]{\mathds{1}_{#1}}

\newcommand{\normnot}[2]{\mathcal{N}\parens{#1,\,#2}}

\newcommand{\uniform}[2]{\mathrm{U}\parens{#1,\,#2}}

\newcommand{\withprob}{~~\text{w.p.}~~}

\newcommand{\errorrv}{\mathcal{E}}

\newcommand{\rstar}{\bv{r}^{\star}}
\newcommand{\yhatstar}{\yhat^{\star}}
\newcommand{\gstar}{g^{\star}}
\newcommand{\hstar}{h^{\star}}